**Short Communication**

# State Space Analysis of Memristor Based Series and Parallel RLCM Circuits

T. D. Dongale [a, ∗], P. K. Gaikwad [b], R. K. Kamat [b]

[a] Computational Electronics and Nanoscience Research Laboratory,
School of Nanoscience and Technology, Shivaji University, Kolhapur 416004, India
[b] Embedded System and VLSI Research Laboratory, Department of Electronics,
Shivaji University, Kolhapur, 416004, India

∗ Corresponding Author: tdd.snst@unishivaji.ac.in

## Abstract

The present paper investigates state space analysis of memristor based series and parallel RLCM circuits. The stability analysis is carried out with the help of eigenvalues formulation method, pole-zero plot and transient response of system. The state space analysis is successfully applied and eigenvalues of the two circuits are calculated. It is found that the, system follows negative real part of eigenvalues. The result clearly shows that addition of memristor in circuits will not alter the stability of system. It is found that systems poles located at left hand side of the $S$ plane, which indicates stable performance of system. It clearly evident that eigenvalues has negative real part hence two systems are internally stable.

**Short Communication**

## I. Introduction:

Memristor is known as fourth basic passive circuit element characterized by its nonlinear relation between charge and magnetic flux and hysteresis loop in I-V plane. The first theoretical concept of memristor was postulated by L. Chua based on symmetric principle with rigorous mathematical proof [1]. The first experimental realization of memristor was confirmed by team of HP group in 2008 [2]. After first realization of memristor, large number of new applications areas are emerged out in last few years. The applications in the memory technology, neuromorphic hardware solution, soft computing, biomedical are name of the few [3-13].

The memristor is well suited for future resistive memory technology due to its high density of data storage, long data retention, low operating voltage, high endurance, fast read and write speed, lower programming current, fast switching speed and compatible with conventional CMOS process [14-15]. The memory property with resistance is a unique kind of feature which leads to many application areas in particular to circuit design. Recently our research group developed nanostructured $TiO_2$ thin film memristor using hydrothermal process with low operation voltage ($\pm 0.68$ V) with high degree of symmetry [16]. In the circuit theory point of view, Driscoll et al reported memristive adaptive memory filter [17]. Talukdar et al reported third-order autonomous linear time variant circuit based on memristor [18]. Pershin et al reported memristor emulator and shows some of the applications of memristors in programmable analog circuit domain [19]. Kolka et al reported frequency domain analysis of circuits with mem-elements and compute steady state analysis using harmonic balance method [20].

The present paper discusses state space analysis for memristor based series and parallel RLCM circuits. The stability analysis is carried out with the help of eigenvalues formulation method, pole-zero plot and transient response of system. In this investigation we have considered memristor as symmetric piecewise-linear driving-point (DP) nonlinear resistor and at a high frequency it behaves like simple linear resistor [21].

## II. Overview of Memristor

The memristor is a simple resistor with memory and it is defined using relationship between charge and magnetic flux depicted in equation (1). It is fundamental and passive element due to fact that the no other combination of resistor, inductor and capacitor mimic its pinched hysteresis loop in I-V plane. The fingerprint characteristics of memristor based on linear drift model are

Short Communication

shown in fig. 1. Memristor does not store any charge itself but it is a totally dissipative circuit element [1]. L. Chua defined memristor based on state variable 'w' and state variable either charge (current controlled memristor) or magnetic flux (voltage controlled memristor) defined in equation (2) and (3) respectively [22].

$$\frac{d\varphi}{dq} = M \qquad (1)$$

$$v = M(q)i \qquad (2)$$

$$i = W(\varphi)v \qquad (3)$$

The $M(q)$ and $W(\varphi)$ are known as memristance and memductance nonlinear functions where, M (q) is in the units of ohms and W ($\varphi$) is in Siemens [22]. The terminal quantities of memristor are given by Pershin et al [15] and Kolka et al [20, 23], based on their formulation,

$$V(t) = \frac{d\varphi\,(q)}{dt} = \frac{d\varphi\,(q)}{dq}\frac{dq}{dt} = R_M(q)\,i\,(t) \qquad (4)$$

$$i(t) = \frac{dq\,(\varphi)}{dt} = \frac{dq\,(\varphi)}{d\varphi}\frac{d\varphi}{dt} = G_M(\varphi)\,V\,(t) \qquad (5)$$

The $R_M\,(q)$ and $G_M\,(\varphi)$ treated as memristance and memductance similar to equation (2) and (3) respectively. For sake of simplicity we concentrate our study only on current controlled memistor.

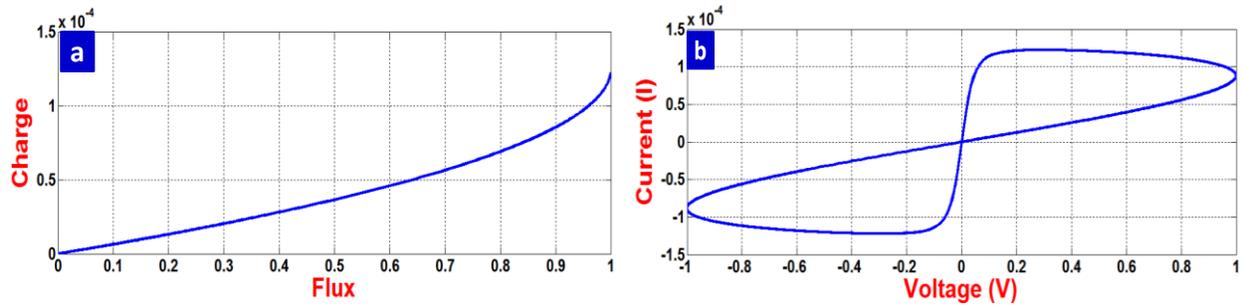

Fig. 1: (a) Plot of Flux V/S Charge. (b) Plot of Current V/S Voltage.

### III. State Space Analysis

*Case I: Series RLCM Circuit*

Fig.2 represents series connection of resistor (R), inductor (L), capacitor (C) and memristor (M) circuit. The circuit is excited by sinusoidal signal *U(t)* and *i* is a total current flowing through the circuit. The $V_L$, $V_R$, Vc, and $V_M$ are the voltage across inductor, resistor, capacitor and memristor

**Short Communication**

respectively. The above circuit consists of two states variable Vc and I<sub>L</sub>, and one output variable Y which are defined by equation (6).

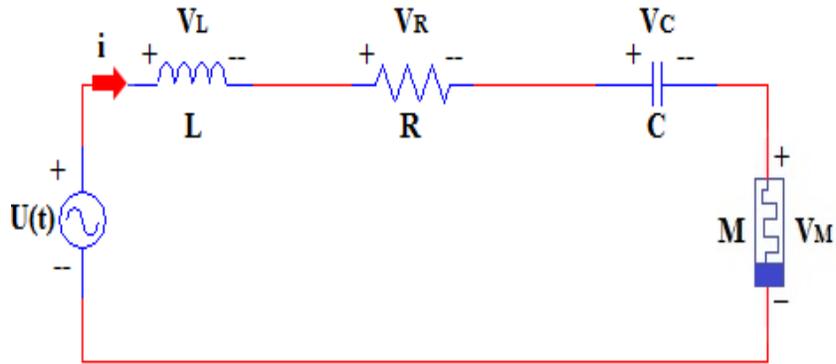

Fig. 2: Series RLCM Circuit

$$\left. \begin{array}{l} X_1 = Vc \\ X_2 = I_L \\ Y = I_L \end{array} \right\} \quad \ldots\ldots\ldots\ldots (6)$$

Applying KVL to above circuit we get,

$$U(t) = V_L + V_R + Vc + V_M$$

$$U(t) = L\frac{dI_L}{dt} + R\, I_L + \frac{dVc}{dt} + R_M(q)\, I_L$$

From above equation we get two relations regarding with state variable,

$$\frac{dVc}{dt} = \frac{I_L}{C} \quad \ldots\ldots\ldots\ldots (7)$$

$$\frac{dI_L}{dt} = \frac{U}{L} - \frac{R}{L} I_L - \frac{1}{L} Vc - \frac{R_M(q)}{L} I_L \quad \ldots\ldots\ldots\ldots (8)$$

Rearranging above equation (7) and (8) we get,

$$\dot{X}_1 = \frac{X_2}{C} \quad \ldots\ldots\ldots\ldots (9)$$

$$\dot{X}_2 = \frac{U}{L} - \frac{R}{L} X_2 - \frac{1}{L} X_1 - \frac{R_M(q)}{L} X_2 \quad \ldots\ldots\ldots\ldots (10)$$

From equation (9) and (10) we can write state matrix in the form of, $\dot{X} = Ax + B_U$ and $Y = Cx$ where, $X = [X_1, X_2]^T$

$$\begin{bmatrix} \dot{X}_1 \\ \dot{X}_2 \end{bmatrix} = \begin{bmatrix} 0 & \frac{1}{C} \\ -\frac{1}{L} & -\left(\frac{R}{L} + \frac{R_M(q)}{L}\right) \end{bmatrix} \cdot \begin{bmatrix} X_1 \\ X_2 \end{bmatrix} + \begin{bmatrix} 0 \\ \frac{1}{L} \end{bmatrix} \cdot U \quad \ldots\ldots\ldots\ldots (11)$$

**Short Communication**

$$Y = \begin{bmatrix} 1 & 1 \end{bmatrix} \cdot \begin{bmatrix} X_1 \\ X_2 \end{bmatrix} \qquad \ldots\ldots\ldots (12)$$

Equation (11) and (12) represent the required state space representation of memristor based series RLCM circuit.

*Case II: Parallel RLCM Circuit*

Fig.3 represents parallel connection of resistor (R), inductor (L), capacitor (C) and memristor (M) circuit. The circuit is excited by sinusoidal signal *Is(t)* and $I_R$, $I_L$, $I_C$, and $I_M$ are the current through resistor, inductor, capacitor, and memristor respectively. The above circuit consists of two states variable Vc and $I_L$ and one output variable Y which are defined as follows,

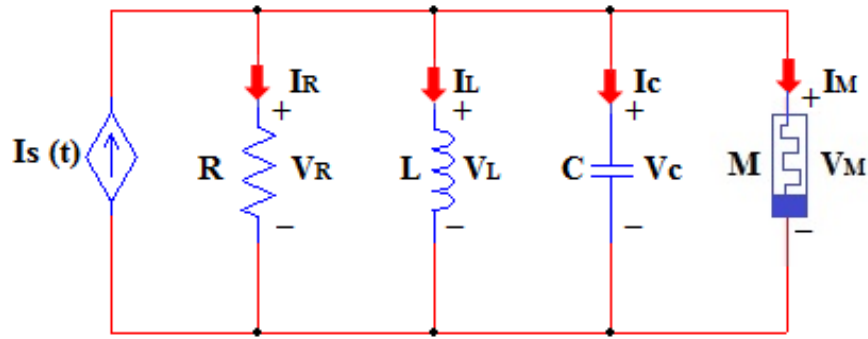

Fig. 3: Parallel RLCM Circuit

$$\left. \begin{array}{l} X_1 = Vc \\ X_2 = I_L \\ Y = Vc \end{array} \right\} \qquad \ldots\ldots\ldots\ldots (13)$$

Applying KCL to above circuit we get,

$$Is(t) = I_R + I_L + I_C + I_M$$

$$Is(t) = \frac{Vc}{R} + I_L + C\frac{dVc}{dt} + \frac{Vc}{R_M(q)}$$

$$L\frac{dI_L}{dt} = Vc$$

From above equation we get two relations regarding with state variable,

$$\frac{dVc}{dt} = -\frac{Vc}{RC} - \frac{I_L}{C} - \frac{Vc}{R_M(q)C} + \frac{Is}{C} \qquad \ldots\ldots\ldots\ldots (14)$$

$$\frac{dI_L}{dt} = \frac{1}{L}Vc \qquad \ldots\ldots\ldots\ldots (15)$$

Rearranging above equation (14) and (15) we get,

$$\dot{X}_1 = -\frac{X_1}{RC} - \frac{X_2}{C} - \frac{X_1}{R_M(q)C} + \frac{Is}{C} \qquad \ldots\ldots\ldots\ldots (16)$$

$$\dot{X}_2 = \frac{1}{L} X_1 \qquad \ldots\ldots\ldots\ldots (17)$$

From equation (16) and (17) we can write state matrix in the form of,

$$\dot{X} = Ax + B_U \text{ and } Y = Cx \text{ where, } X = [X_1, X_2]^T$$

$$\begin{bmatrix} \dot{X}_1 \\ \dot{X}_2 \end{bmatrix} = \begin{bmatrix} -\left(\frac{1}{RC} + \frac{1}{R_M(q)C}\right) & -\frac{1}{C} \\ \frac{1}{L} & 0 \end{bmatrix} \cdot \begin{bmatrix} X_1 \\ X_2 \end{bmatrix} + \begin{bmatrix} \frac{1}{C} \\ 0 \end{bmatrix} \cdot I_S \qquad \ldots\ldots\ldots (18)$$

$$Y = \begin{bmatrix} 1 & 1 \end{bmatrix} \cdot \begin{bmatrix} X_1 \\ X_2 \end{bmatrix} \qquad \ldots\ldots\ldots (19)$$

Equation (18) and (19) represent the required state space representation of memristor based Parallel RLCM circuit.

## IV. Eigenvalues of Series and Parallel RLCM Circuits

The eigenvalues of memristor based series and parallel RLCM circuit are as follows,

$$\lambda_{1,2 \text{ (Series)}} = -\left(\frac{R+R_M(q)}{2}\right) \pm \sqrt{\frac{C(R+R_M(q))^2 - 4}{LC}} \qquad \ldots\ldots (20)$$

$$\lambda_{1,2 \text{ (Parallel)}} = -\left(\frac{\left(\frac{1}{RC}+\frac{1}{R_M(q)C}\right) \pm \sqrt{\left(\frac{1}{RC}+\frac{1}{R_M(q)C}\right)^2 - 4\left(\frac{1}{LC}\right)}}{2}\right) \qquad \ldots\ldots\ldots (21)$$

The equation (20 and 21) represent eigenvalues of the memristor based series and parallel circuit respectively. It is clearly evident that eigenvalues have negative real part and hence the above two systems are internally stable.

## V. Time Domain Analysis of Series and Parallel RLCM Circuits

The pole-zero plot and system response to standard test signal (step and impulse) are derived to define stability of system. The system transient response is depicted in fig. 4 (a-b) and pole-zero plots of series and parallel RLCM circuit are shown in fig. 4 (c-d) respectively. The performance parameter of both the circuits is shown in table 1. It is clearly evident from fig. 4 (a-b) that transient responses of systems possess bounded input bounded output (BIBO) stability. The result clearly indicates series RLCM circuit having complex root with negative real parts and parallel RLCM circuit having distinct root with negative real parts. The overall result shows BIBO stability of two systems.

**Short Communication**

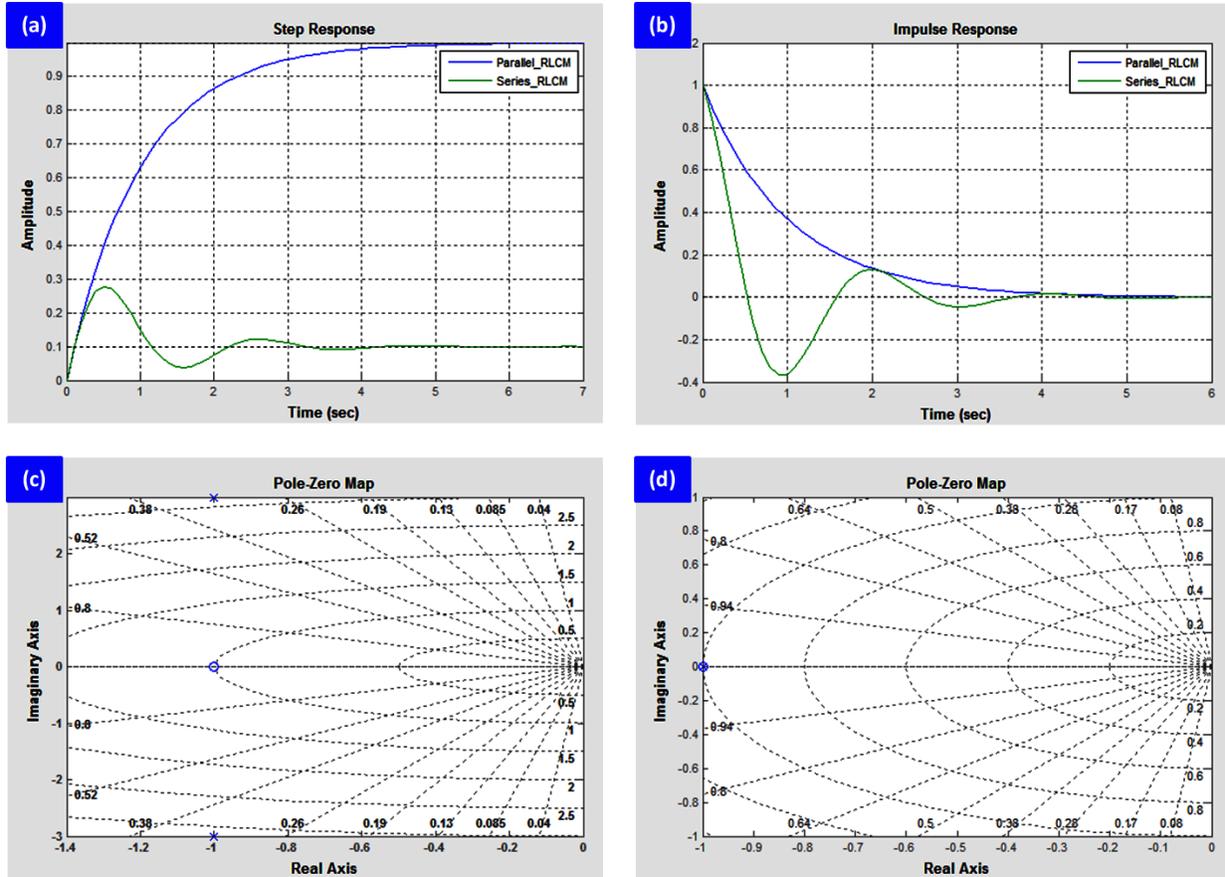

Fig. 4: (a) Step Response of Series and parallel RLCM Circuit. (b): Impulse Response of Series and Parallel RLCM Circuit. (c): Pole-Zero map of Series RLCM Circuit. (d): Pole-Zero map of parallel RLCM Circuit.

Table I: Performance Parameter of Series and Parallel RLCM Circuit

| Performance Parameter | | Series RLCM | Parallel RLCM |
|---|---|---|---|
| Peak Amplitude | | 0.27 | 0.99 |
| Overshoot (%) | | 177 | 0.00 |
| Peak Time (Second) | | 0.49 | 6.00 |
| Rise Time (Second) | | 0.08 | 2.20 |
| Settling Time (Second) | Step Response | 4.06 | 3.91 |
| Final Value | | 0.10 | 1.00 |
| Pole Location | | $\alpha_1 = -1+3j$ and $\alpha_2 = -1-3j$ | Absent |
| Zero Location | | Z= -1 | Z= -1 |
| Peak Amplitude | Impulse Response | 1.00 | 1.00 |
| Settling Time (Second) | | 3.44 | 3.91 |

**Short Communication**

## VI. Conclusion

The present paper deals with the state space analysis of memristor based series and parallel circuit. The state space numerical analysis with eigenvalues methodology is adopted for stability investigation. The eigenvalues of system has negative real part so the above two system are internally stable. It simply means that the pole of system located at left hand side of 'S' plane. The system responses to standard test signal (step and impulse) also shows the stability of systems. The results clearly indicate that, addition of memristor in circuit will not alter the performance of system.

**Short Communication**